\documentstyle[aps,prd,twocolumn,epsfig]{revtex}

\title{Gapless two-flavor color superconductor}

\author{Igor Shovkovy$^{1,*}$ and Mei Huang$^{1,2}$}

\address{$^1$Institut f\"{u}r Theoretische Physik, 
J.W. Goethe-Universit\"{a}t, D-60054 Frankurt/Main, Germany}

\address{$^2$Physics Department, Tsinghua University, 
Beijing 100084, China}

\date{\today}

\draft

\begin{document} 
\maketitle

\begin{abstract}

A new, gapless two-flavor color superconducting phase that appears 
under conditions of local charge neutrality and $\beta$-equilibrium 
is revealed. In this phase, the symmetry of the ground state is the 
same as in the conventional two-flavor color superconductor. In the 
low-energy spectrum of this phase, however, there are only two gapped
fermionic quasiparticles, and the other four quasiparticles are
gapless. The origin and the basic properties of the gapless two-flavor
color superconductor are discussed. This phase is a natural candidate 
for quark matter in cores of compact stars.

\end{abstract}


Sufficiently cold and dense quark matter is a color superconductor
\cite{old,cs}. The ground state is characterized by a condensate of
Cooper pairs that are made of quarks with equal and opposite momenta. In
this phase of matter, the SU(3)$_{c}$ color gauge group is broken
(partially or completely) through the Anderson-Higgs mechanism. At
asymptotic densities, this phenomenon was studied in detail from first
principles in Refs. \cite{weak,PR-sp1,weak-cfl}.

In general, quark matter at high baryon density has a rich phase
structure, consisting of different normal and color-superconducting
phases. Changing the baryon chemical potential, it should be possible to
obtain phases in which either two or three of the lightest quark flavors
participate in Cooper pairing.

At not very large densities, one gets quark matter made of only up and
down quarks. If the Fermi momenta of different types of quarks are
approximately the same, the up and down quarks will form Cooper pairs in
the color-antitriplet, flavor-singlet, spin-zero channel
\cite{cs,weak,PR-sp1}. The corresponding ground state of matter is the
so-called two-flavor color superconductor (2SC). In this phase, the color
gauge group is broken by the Anderson-Higgs mechanism down to SU(2)$_{c}$
subgroup. With the conventional choice of the condensate pointing in the
``blue" direction, one finds that the condensate consists of red up
($u_{r}$) and green down ($d_{g}$), as well as green up ($u_{g}$) and red
down ($d_{r}$) diquark Cooper pairs. The other two quarks ($u_{b}$ and
$d_{b}$) do not participate in pairing. This is the conventional picture
of the 2SC phase \cite{cs,weak,PR-sp1}.

At sufficiently large densities, i.e., when the value of the chemical
potential exceeds the constituent (medium modified) mass of the strange
quark, quark matter should consist of all three quark flavors. This
opens the possibility for Cooper pairing between the three lightest
quarks (up, down and strange). Depending on the value of the strange
quark mass, as well as other parameters in the theory, one might get
the color-flavor locked phase (CFL) \cite{cfl}, or even the more exotic
crystalline phase \cite{crystal} of quark matter. Besides that, it is
also possible that up and down quarks form 2SC matter, while the strange
quarks do not participate in pairing. We call this latter phase 2SC+s.
Still another possibility is that the strange quarks by themselves form a
color superconducting condensate. This will be then a spin-1 condensate
\cite{PR-sp1,spin-1}.

It is natural to expect that some color superconducting phases may exist
in the interior of compact stars. The estimated central densities of such
stars might be as large as $10 \rho_{0}$ (where $\rho_{0}\approx 0.15$
fm$^{-3}$ is the saturation density), while their temperatures should be
in the range of tens of keV. These values are very encouraging.

Matter in the bulk of compact stars should be neutral with respect to
electric as well as color charges. Also, such matter should remain in
$\beta$-equilibrium. Satisfying these requirements may impose nontrivial
relations between the chemical potentials of different quarks. In turn,
such relations could substantially influence the pairing dynamics between
quarks, for instance, by suppressing some color superconducting phases
and by favoring others.

It was argued in Ref. \cite{absence2sc}, that the 2SC+s phase becomes
less favorable than the CFL phase if the charge neutrality condition is
enforced (note that the strange quark mass was chosen too small to allow
the appearance of a pure 2SC phase in Ref. \cite{absence2sc}). Therefore,
one might speculate that only CFL quark matter could exist inside compact
stars. In such a case, the outside hadronic layer of the star would make
a direct contact with the CFL quark core through a separating sharp
interface \cite{interface,AlforReddy}. The baryonic density and the
energy density would have a jump (smoothed only over microscopic
distances) at the interface. Of course, a mixed phase of hadronic and CFL
matter is another possibility, but it does not seem to be favorable
\cite{AlforReddy} (for other studies of the stars with CFL quark matter
in their interior see Ref. \cite{compact_CFL}).

We should mention that the general conclusion of Ref. \cite{absence2sc}
was essentially confirmed in Ref. \cite{neutral_steiner} where it was
claimed that the 2SC+s could exist only in a narrow window (about $10$ to
$15$ MeV wide) of baryon chemical potential around the midpoint $\mu
\equiv \mu_{B}/3\approx 450$ MeV. For lower values of $\mu$, no neutral
2SC phase was found. In this Letter, however, we argue that a neutral 2SC
phase does exist at lower values of $\mu$. It is a {\em gapless} rather
than an ordinary 2SC phase. In particular, the conventional density
relations between the pairing quarks, used in Ref.
\cite{neutral_steiner}, are not valid in this new type of 2SC quark
matter. This phase is rather unusual because it has the same symmetry of
the ground state as the conventional 2SC phase, but the spectrum of its
fermionic quasiparticles is very different.

It appears that the neutral gapless 2SC phase of quark matter has already
been used in numerical studies of Refs. \cite{huang_2sc,Ruester} (and,
possibly, in Ref. \cite{Blaschke_2sc}). Its special properties, however,
have not been appreciated and have never been discussed before. In this
Letter, we are going to explain the origin of this new phase and shed
some light on its properties.

Let us start our analysis from discussing the quark model that we use.
Without loosing generality, we assume that the strange quark is
sufficiently heavy and does not appear at intermediate baryon densities
under consideration (e.g., this might correspond to the baryon chemical
potential $\mu\equiv\mu_{B}/3$ in the range between about $350$ and $450$
MeV). Then, in our study, we could use the simplest SU(2)
Nambu--Jona-Lasinio (NJL) model \cite{huang_2sc}. The explicit form of
the Lagrangian density reads:
\begin{eqnarray}
\label{lagr}
{\cal L} & = &\bar{q}(i\gamma^{\mu}\partial_{\mu}-m_0)q + 
 G_S\left[(\bar{q}q)^2 + (\bar{q}i\gamma_5{\bf \vec{\tau}}q)^2\right] 
\nonumber \\
 &&+G_D\left[(i \bar{q}^C  \varepsilon  \epsilon^{b} \gamma_5 q )
   (i \bar{q} \varepsilon \epsilon^{b} \gamma_5 q^C)\right],
\end{eqnarray}
where $q^C=C {\bar q}^T$ is the charge-conjugate spinor and $C=i\gamma^2
\gamma^0$ is the charge conjugation matrix. The quark field $q \equiv
q_{i\alpha}$ is a four-component Dirac spinor that carries flavor
($i=1,2$) and color ($\alpha=1,2,3$) indices. ${\vec \tau} =(\tau^1,
\tau^2, \tau^3)$ are Pauli matrices in the flavor space, while
$(\varepsilon)^{ik} \equiv \varepsilon^{ik}$ and $(\epsilon^b)^{\alpha
\beta} \equiv \epsilon^{\alpha \beta b}$ are antisymmetric tensors in
flavor and color, respectively. We also introduce a momentum cut-off
$\Lambda$, and two independent coupling constants in the scalar
quark-antiquark and scalar diquark channels, $G_S$ and $G_D$.

The values of the parameters in the NJL model are chosen as follows:
$G_S=5.0163$ GeV$^{-2}$ and $\Lambda=0.6533$ GeV \cite{huang_2sc}. In
this Letter, we consider only the chiral limit with $m_0=0$. As for the
strength of the diquark coupling $G_D$, its value is taken to be
proportional to the quark-antiquark coupling constant, i.e., $G_D = \eta
G_S$ with a typical number for $\eta$ being around $0.75$
\cite{neutral_steiner,huang_2sc}.

In $\beta$-equilibrium, the diagonal matrix of quark chemical potentials
is given in terms of baryonic, electric and color chemical potentials,
\begin{equation} 
\mu_{ij, \alpha\beta}= (\mu \delta_{ij}- \mu_e Q_{ij})
\delta_{\alpha\beta} + \frac{2}{\sqrt{3}}\mu_{8} \delta_{ij} 
(T_{8})_{\alpha \beta},
\end{equation} 
where $Q$ and $T_8$ are generators of U(1)$_{em}$ of electromagnetism
and the U(1)$_{8}$ subgroup of the color gauge group. The explicit
expressions for the quark chemical potentials read
\begin{eqnarray}
\mu_{ur} =\mu_{ug} =\mu -\frac{2}{3}\mu_{e} +\frac{1}{3}\mu_{8}, \\
\mu_{dr} =\mu_{dg} =\mu +\frac{1}{3}\mu_{e} +\frac{1}{3}\mu_{8}, \\
\mu_{ub} =\mu -\frac{2}{3}\mu_{e} -\frac{2}{3}\mu_{8}, \\
\mu_{db} =\mu +\frac{1}{3}\mu_{e} -\frac{2}{3}\mu_{8}. 
\end{eqnarray}

The effective potential for quark matter at zero temperature and
in $\beta$-equilibrium with electrons takes the form
\begin{equation} 
\label{potential}
\Omega = \Omega_{0}-\frac{\mu_e^4}{12 \pi^2}
+\frac{m^2}{4G_S}+\frac{\Delta^2}{4G_D}
- \sum_{a} \int\frac{d^3 p}{(2\pi)^3} |E_{a}|,
\end{equation}   
where $\Omega_{0}$ is a constant added to make the pressure of the vacuum
zero. This is the zero temperature limit of the potential derived in Ref. 
\cite{huang_2sc}. The sum in Eq.~(\ref{potential}) runs over all (6 quark
and 6 antiquark) quasiparticles. The dispersion relations and the degeneracy
factors of the quasiparticles read
\begin{eqnarray} 
E_{ub}^{\pm} &=& E(p) \pm \mu_{ub} , \hspace{26.6mm} [\times 1]
\label{disp-ub} \\
E_{db}^{\pm} &=& E(p) \pm \mu_{db} , \hspace{26.8mm} [\times 1]
\label{disp-db}\\
E_{\Delta^{\pm}}^{\pm} &=& \sqrt{[E(p) \pm \bar{\mu}]^2
+\Delta^2} \pm  \delta \mu ,\hspace{6mm} [\times 2]
\label{2-degenerate}
\end{eqnarray}
where the shorthand notations are $E(p)\equiv \sqrt{{\bf p}^2+m^2}$,
$\bar{\mu}\equiv (\mu_{ur} +\mu_{dg})/2 =\mu-\mu_{e}/6+\mu_{8}/3$, and
$\delta \mu\equiv (\mu_{dg}-\mu_{ur})/2 =\mu_{e}/2$.

\begin{figure}
\epsfxsize=8cm
\epsffile[88 0 588 313]{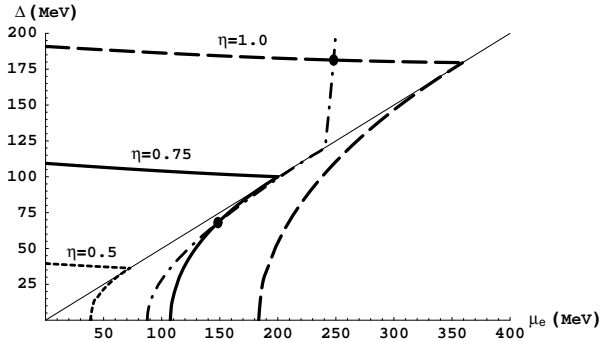}

\caption{The graphical representation of the solutions to the gap equation
for three different values of the diquark coupling constant (thick solid
and dashed lines), and to the electrical neutrality condition (thick
dash-dotted line). The intersection points represent the solutions to both
equations. The thin solid line divides two qualitatively different
regions, $\Delta<\mu_e/2$ and $\Delta>\mu_e/2$. The results are plotted
for $\mu=400$ MeV and three values of diquark coupling constant $G_{D} =
\eta G_S$ with $\eta=0.5$, $\eta=0.75$, and $\eta=1.0$.}
 
\label{gapneutral}
\end{figure}

In this study we concentrate exclusively on the 2SC phase of quark
matter. The numerical study shows that the constituent quark mass $m$ is
zero in this phase \cite{huang_2sc}. Then by making use of the
dispersion relations, we derive
\begin{eqnarray}
\Omega &=& \Omega_{0}
-\frac{\mu_e^4}{12 \pi^2} 
+\frac{\Delta^2}{4G_D} 
-\frac{\Lambda^4}{2\pi^2} 
-\frac{\mu_{ub}^4}{12 \pi^2}
-\frac{\mu_{db}^4}{12 \pi^2} \nonumber \\
&-& 2 \int_{0}^{\Lambda} \frac{p^2 d p}{\pi^2} 
\left(\sqrt{(p+ \bar{\mu})^2+\Delta^2}
+\sqrt{(p-\bar{\mu})^2+\Delta^2}\right)\nonumber \\
&-& 2\theta \left(\delta\mu-\Delta\right)
\int_{\mu^{-}}^{\mu^{+}}\frac{p^2 d p}{\pi^2}\Big(
\delta\mu-\sqrt{(p-\bar{\mu})^2+\Delta^2}
\Big),
\label{pot-2sc}
\end{eqnarray}
where $\mu^{\pm}\equiv \bar{\mu}\pm \sqrt{(\delta\mu)^2-\Delta^2}$. Note
that the physical thermodynamic potential that determines the pressure,
$\Omega_{\rm phys} =-P$, is obtained from $\Omega$ in Eq. (\ref{pot-2sc})
after substituting $\mu_{8}$ , $\mu_{e}$ and $\Delta$ that solve the
color and electrical charge neutrality conditions, as well as the gap
equation, i.e.,
\begin{equation} 
n_{8}\equiv
\frac{\partial \Omega}{\partial \mu_{8}}=0, \quad n_{Q}\equiv
\frac{\partial \Omega}{\partial \mu_{e}}=0, \quad \mbox{and} \quad
\frac{\partial \Omega}{\partial \Delta}=0.
\label{3conditions}
\end{equation} 
Let us assume that the solution to the color neutrality condition,
$\mu_{8}(\mu_{e},\Delta)$, is known for a given value of $\mu$. In
practice, we construct this function using a numerical set of solutions.
By substituting this solution into the potential in Eq. (\ref{pot-2sc}),
one is left with a function $\Omega$ that depends only on two parameters,
$\mu_{e}$ and $\Delta$. In this two-dimensional parameter space, the
solution to each of the extra two conditions in Eq. (\ref{3conditions})
is a one-dimensional line. Graphically, this is shown in Fig.
\ref{gapneutral}. Since both conditions should be satisfied
simultaneously, one needs to find the intersection points of the
corresponding lines.

First we discuss the neutrality condition, $n_{Q}=0$. The corresponding
line of solutions in the two-dimensional parameter space is a monotonic,
single valued function. In Fig. \ref{gapneutral}, it is represented by 
the thick dash-dotted line. This line of solutions consists of two
branches in two qualitatively different regions of the parameter space
that are separated by the line $\Delta=\mu_e/2$. In the leading order
approximation, this neutrality line is independent of the coupling
constant $G_{D}$. This should be obvious from the derivation of the first
two conditions in Eq. (\ref{3conditions}).

The line of solutions to the gap equation is most interesting. It is made
of two different branches in two regions, $\Delta<\mu_e/2$ and
$\Delta>\mu_e/2$, see Fig. \ref{gapneutral}. The upper is the main branch
that exists down to $\mu_{e}=0$ (i.e., no mismatch in the Fermi surfaces
of the up and down quarks). The lower branch appears only in a finite
window of electric chemical potentials, and it merges with the upper one
at a point on the line $\Delta=\mu_e/2$. In Fig. \ref{gapneutral}, we show
three solutions to the gap equation at $\mu=400$ MeV in three regimes with
different coupling constants, corresponding to $\eta=0.5$, $\eta=0.75$,
and $\eta=1.0$. [The results look the same at all values of $\mu$.] The
three curves have qualitatively the same shape, but differ by overall
scale factors. The difference in the overall scale has an important
consequence. At weak coupling ($\eta\alt0.7$), there is no neutral 2SC
phase because there is no intersection of the solution to the gap equation
with the neutrality line, see $\eta=0.5$ solution in Fig.
\ref{gapneutral}. At intermediate ($0.7 \alt \eta \alt 0.8$) and strong
($\eta \agt 0.8$) coupling, on the other hand, there is an intersection
with the neutrality line at a point on the lower and upper branches,
respectively. In these last two cases, neutral 2SC phases exist.

If the electric chemical potential were a free parameter, one would find
that increasing its value leads to a first order phase transition
\cite{bedaque}. This is the result of the appearance of two competing
local minima (at $\Delta =0$ and $\Delta \neq 0$)  of the potential
$\Omega(\Delta , \mu_e=\mbox{const})$ for a range of intermediate values
of the parameter $\mu_e$. Note, that the location of the local maximum of
the potential that lies between the two local minima is determined by the
lower ``unstable" branch of the solutions to the gap equation in
Fig.~\ref{gapneutral}. The first order phase transition happens at about
$\mu^{\rm cr}_{e}= \sqrt{2} \Delta_{0}$ where $\Delta_{0}$ is the value of
the gap at $\mu_{e} =0$ (this estimate is derived in the approximation of
weak coupling \cite{mag-crit}). From our discussion below, it will be
clear that this first order phase transition is unphysical under the
requirement of local neutrality of quark matter. This is because the
transition typically happens between two types of matter with different,
{\em nonzero} charge densities (e.g., positively charged color
superconducting matter and negatively charged normal quark matter). It is
worth mentioning, however, that this first order phase transition can get
physical meaning in a globally neutral mixed phase of quark matter
\cite{SHH}, and in some condensed matter systems where there is no
analogue of the charge neutrality condition \cite{sarma}.

\begin{figure}
\epsfxsize=8cm
\epsffile{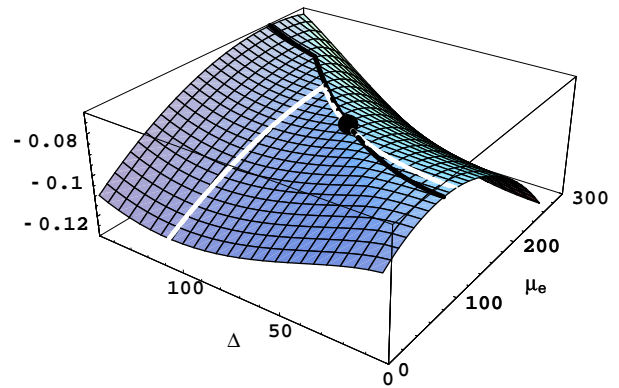}

\caption{The effective potential as a function of the diquark gap
$\Delta$ and electric chemical potential $\mu_e$. The black solid line
gives the potential along the electric neutrality line. The white solid
line shows the line of solutions to the gap equation. The results are
plotted for $\mu=400$ MeV and $G_{D}=\eta G_{S}$ with $\eta=0.75$.}

\label{V3D}   
\end{figure}

The three-dimensional view of the potential $\Omega$ as a function of two
parameters, $\Delta$ and $\mu_{e}$, is represented by a surface in Fig.
\ref{V3D}. The results are plotted for $\mu=400$ MeV and for the diquark
coupling with $\eta=0.75$. In this study, we are interested exclusively in
locally neutral quark matter (mixed phases are discussed in Ref.
\cite{SHH}), and therefore we consider the potential only along the
neutrality line (represented by a black solid line in Fig. \ref{V3D}).

\begin{figure}
\epsfxsize=8cm
\epsffile{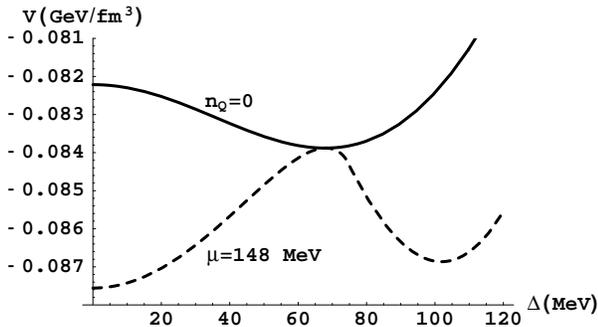}

\caption{The effective potential as a function of the diquark gap $\Delta$
calculated at a fixed value of the electric chemical potential $\mu_e =
148.4$ MeV (dashed line), and the effective potential defined along the
neutrality line (solid line). The results are plotted for $\mu=400$ MeV
and $G_{D}=\eta G_{S}$ with $\eta=0.75$.}

\label{V2D}
\end{figure}

The study shows that, for any coupling strength, such a potential of the
neutral quark matter, $\Omega(\Delta,\mu_e|_{n_Q=0})$, as a function of
$\Delta$ has one local minimum which is also the global one. In the regime
of intermediate strength of diquark coupling (e.g., $\eta=0.75$), this
minimum corresponds to a point on the lower branch of the solutions to the
gap equation in Fig.~\ref{gapneutral} (i.e., $\Delta \approx 68$ MeV and
$\mu_e \approx 148.4$ MeV). This may sound very surprizing because the
points on the lower branch correspond to local maxima of the potential
$V(\Delta, \mu_e=\mbox{const})$. To make this clear, we compare the
effective potential calculated for a fixed value of $\mu_e$ with the
potential defined along the neutrality line in Fig.~\ref{V2D}. As we see,
after enforcing the condition of charge neutrality, a local maximum of the
former becomes a global minimum of the latter.

It is interesting to study the dependence of the diquark coupling constant
on the properties of the potential of neutral matter. In the weakly
coupled regime ($\eta \alt 0.7$), the potential has the global minimum at
$\Delta=0$. This is in agreement with the fact that the neutrality line
does not intersect the corresponding line of nontrivial solutions (e.g.,
$\eta=0.5$ case in Fig. \ref{gapneutral}). For intermediate values of the
coupling ($0.7 \alt \eta \alt 0.8$), the extremum at $\Delta=0$ is a local
maximum, while a minimum appears at $\Delta\neq 0$, see solid curve in
Fig. \ref{V2D}. In this case, the minimum corresponds to an intersection
of the neutrality line with the lower branch of the solutions to the gap
equation (e.g., $\eta=0.75$ case in Fig. \ref{gapneutral}). Finally, in
the strong coupling regime ($\eta \agt 0.8$), there is a nontrivial
minimum which corresponds to an intersection of the neutrality line with
the upper branch of the solutions (e.g., $\eta=1.0$ case in Fig.
\ref{gapneutral}). The phase transition controlled by the coupling constant
$\eta$ is a second order phase transition.

\begin{figure}
\epsfxsize=8cm
\epsffile[88 -10 488 252]{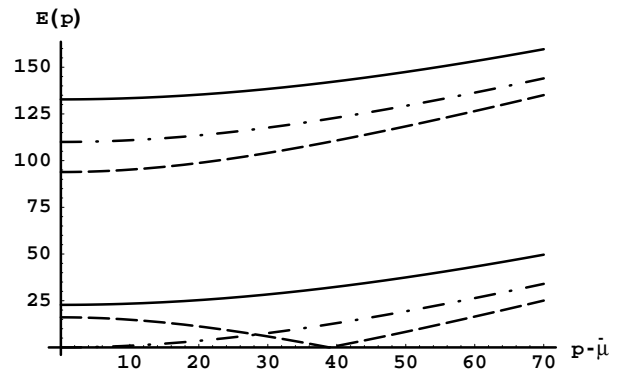}

\caption{Three types of dispersion relations of quasiparticles that
originate from quarks of two (red and green) colors participating in
Cooper pairing. The two solid (dashed) lines represent the dispersion
relations in the regime with $\Delta>\mu_{e}/2$ ($\Delta<\mu_{e}/2$). The
dash-dotted line corresponds to the special case with $\Delta=\mu_{e}/2$.
Each line corresponds to two degenerate quasiparticles.}

\label{dispersion}
\end{figure}

Here we argue that the regime of intermediate couplings describes a new,
gapless phase of 2SC quark matter. This phase of superconducting quark
matter possesses {\em four} gapless and only two gapped fermionic
quasiparticles in its spectrum, see Eqs.
(\ref{disp-ub})--(\ref{2-degenerate}). Recall that the ordinary 2SC phase
(which, in fact, appears in the region $\Delta>\mu_{e}/2$ in the strong
coupling regime) has only {\em two} gapless and four gapped
quasiparticles. The number of gapless modes changes when one crosses the
line $\Delta=\mu_{e}/2$. The quasiparticle dispersion relations,
originating from the red and green quarks, as described by Eq.
(\ref{2-degenerate}), are shown in Fig. \ref{dispersion}. Each line
corresponds to two degenerate quasiparticles. The dispersion relations of
gapless blue quarks, given in Eqs. (\ref{disp-ub}) and (\ref{disp-db}),
are not shown in Fig. \ref{dispersion}.

In ordinary 2SC matter (with a solution in the region $\Delta >
\mu_{e}/2$), there are two doublets of gapped modes: one with the gap
$\Delta-\mu_{e}/2$ and the other with the gap $\Delta+\mu_{e}/2$. As one
approaches the boundary between the regions, $\Delta\to\mu_{e}/2$, the
two quasiparticles of the first doublet with the smaller gap gradually
become gapless. They also remain gapless in the phase with
$\Delta<\mu_{e}/2$ (regime of intermediate coupling). This justifies the
name ``gapless 2SC phase". 

In the study of the CFL phase in Ref. \cite{enforced}, it was revealed
that the number densities of the pairing quarks are equal. By applying
the same arguments in the case of 2SC quark matter, one might expect to
get similar relations for the number densities of red up and green down
(as well as green up and red down) quarks. Moreover, these relations have
been numerically confirmed for the 2SC phase of quark matter in the
strongly coupled regime in Ref. \cite{neutral_buballa}. We find, however,
that the number densities of pairing quarks are not always equal. In
particular, they are not equal in the gapless 2SC phase, i.e.,
$n_{ur}=n_{ug} \neq n_{dr}=n_{dg}$. This is directly related to the fact
that the corresponding solution is found in the region
$\Delta<\mu_{e}/2$, where two extra gapless modes appear.

In conclusion, the analysis of this Letter confirms that the charge
neutrality condition can strongly affect color superconducting quark
matter, favoring some phases and disfavoring others. Moreover, as we
argued, by imposing the charge neutrality condition, one can also obtain
new stable phases of matter which could not exist otherwise. The gapless
2SC phase of quark matter is a beautiful example of such a new phase. The
symmetry of its ground state is the same as that of the conventional 2SC
phase. However, the energy spectrum of the fermionic quasiparticles in
gapless 2SC matter is different: it has two additional gapless modes. We
should note that this phase is very different from another example of
gapless color superconductivity, discussed in Ref. \cite{gaplessCFL},
which is a metastable state of CFL quark matter. The gapless 2SC phase is
a stable, neutral state of two-flavor quark matter.

In nature, gapless 2SC quark matter could exist in compact stars. Indeed,
this phase is neutral with respect to electric and color charges and
satisfies the $\beta$-equilibrium condition by construction. Of course,
satisfying these requirements is necessary, but not sufficient. In order
to decide whether the gapless 2SC phase is likely to appear inside stars,
further detailed studies are needed.

We could speculate that the thermodynamic properties of the gapless 2SC
phase should be closer to the properties of neutral normal quark matter
rather than to those of strongly coupled 2SC matter. Qualitatively, this
is expected from observing the relative shallowness of the effective
potential in Fig. \ref{V2D} defined along the neutrality line. It
indicates that, for a given value of the baryon chemical potential $\mu$,
the pressure difference of (neutral) normal and gapless 2SC quark matter
is substantially smaller than the pressure difference of the normal quark
and conventional 2SC phase, $\delta P=(\bar{\mu}\Delta/\pi)^{2}$
\cite{enforced,us-pressure}.

Here we studied only the most general properties of the gapless 2SC phase
of quark matter. By taking into account that this phase is a realistic
candidate for matter inside compact star cores, it would be very
interesting to study its other properties that could potentially affect
some compact star observables. In particular, this includes magnetic
properties, neutrino emissivities, and various transport properties of
gapless 2SC matter.

{\bf Acknowledgements}.  The authors thank Prof. D.H.~Rischke for
stimulating discussions that triggered this study, as well as M.~Hanauske
and S.~R\"uster for useful comments and discussions. We would like to
thank Prof.  H. St\"{o}cker for his kind hospitality at the Institut
f\"{u}r Theoretische Physik. M.H. acknowledges the financial support from
Bundesministerium f\"{u}r Bildung und Forschung (BMBF), the Alexander von
Humboldt-Foundation, and the NSFC under Grant Nos.  10105005, 10135030.
The work of I.A.S. was supported by Gesellschaft f\"{u}r
Schwerionenforschung (GSI) and by Bundesministerium f\"{u}r Bildung und
Forschung (BMBF).

\end{document}